\documentclass{ws-mpla}

\newcommand{\Lie}{\pounds}
\def\dual{{}^\star\!}

\begin{document}

\markboth{Chen, Liu, Nester} {Quasi-local energy for cosmological
models}

\catchline{}{}{}{}{}

\title{Quasi-local energy for cosmological models}

\author{\footnotesize Chiang-Mei Chen}
\address{Department of Physics, National Central University,
Chungli, 320 Taiwan, ROC
\\
cmchen@phy.ncu.edu.tw}

\author{Jian-Liang Liu}
\address{Department of Physics, National Central University,
Chungli, 320 Taiwan, ROC}

\author{James M. Nester}
\address{Department of Physics and Institute of Astronomy,
\\
National Central University, Chungli, 320 Taiwan, ROC
\\
nester@phy.ncu.edu.tw}

\maketitle

\pub{Received (Day Month Year)}{Revised (Day Month Year)}

\begin{abstract}
First we briefly review our covariant Hamiltonian approach to
quasi-local energy, noting that the Hamiltonian-boundary-term
quasi-local energy expressions depend on the chosen boundary
conditions and reference configuration. Then we present the
quasi-local energy values resulting from the formalism applied to
homogeneous Bianchi cosmologies. Finally we consider the quasi-local
energies of the FRW cosmologies. Our results do not agree with
certain widely accepted quasi-local criteria. \keywords{quasi-local
energy; Hamiltonian; cosmology.}
\end{abstract}


\ccode{PACS Nos.: 04.20.Cv, 04.20.Fy, 98.80.Jk}

\section{Introduction}
The localization of energy-momentum for any gravitating system (and
thus for all physical systems) is still an outstanding fundamental
problem. In view of conservation, and the fact that sources exchange
energy-momentum locally with the gravitational field, some kind of
local description for gravitational energy-momentum was expected.
However all attempts at constructing such an expression led only to
reference frame dependent quantities, generally referred to as
pseudotensors.\cite{CNC99PRL,N04CQG} It became apparent that the
gravitational field itself, unlike all matter and other interaction
fields, has no proper energy-momentum density. This fact can be
understood as a consequence of Einstein's equivalence
principle.\cite{MTW} The energy-momentum of gravity---and thus the
energy-momentum for all physical systems---is inherently {\it
non-local}. The modern idea is {\it quasi-local\/}: energy-momentum
is associated with a closed surface bounding a region.

Many quasi-local expressions have been proposed, but presently there
is no consensus as to which is the most suitable, or even as to
which properties a good expression should have.\cite{S04LRR}
Various lists of desiderata for a ``physical'' quasi-local energy
have been presented; according to a well-known one\cite{LY04} the
quasi-local energy should be
\begin{itemize}
\item
zero for flat space,

\item
for spherical symmetric $\simeq$ standard value,

\item
the ADM mass at spatial infinity,

\item
the Bondi mass at null infinity,

\item
for the apparent horizon $\simeq$ standard value,

\item
positive.
\end{itemize}
Our Hamiltonian based quasi-local results do not satisfy the first
and last of these criteria. There is a stronger form\cite{S04LRR} of
the first requirement, namely that the energy vanish iff the
quasilocal region is flat space. Our analysis of the quasi-local
energy of cosmological regions leads us to propose a certain
modification of this stronger form.

\section{The covariant Hamiltonian approach}
Here we briefly summarize the relevant parts of our covariant
Hamiltonian approach to quasi-local
energy.\cite{N04CQG,CNT95PLA,CN99,CNC99,CN00GC,CNT05PRD}

\subsection{First order Lagrangian}
The first order Lagrangian for an f-form field $\varphi$ and its
conjugate momentum $p$ has the form
\begin{equation}
{\cal L} = d \varphi \wedge p - \Lambda(\varphi, p).
\end{equation}
The variation of this 4-form,
\begin{equation}
\delta {\cal L} = d(\delta \varphi \wedge p) + \delta \varphi \wedge
\frac{\delta {\cal L}}{\delta \varphi} + \frac{\delta {\cal
L}}{\delta p} \wedge \delta p, \label{varL}
\end{equation}
leads to the first order equations of motion
\begin{equation}
\frac{\delta {\cal L}}{\delta p} := d \varphi - \partial_p \Lambda =
0, \quad \frac{\delta {\cal L}}{\delta \varphi} := - \varsigma d p -
\partial_\varphi \Lambda = 0,
\end{equation}
where $\varsigma := (-1)^f$.

As a simple example of this formalism consider electromagnetism. The
first order Lagrangian 4-form for the (source free) U(1) gauge field
one-form $A$ is
\begin{equation}
{\cal L}_{\rm EM} = d A \wedge H - \frac12 \dual H \wedge H.
\end{equation}
Variation leads to the pair of first order equations
\begin{equation}
d H = 0, \qquad d A - \dual H = 0.
\end{equation}
From $F: = dA= \dual H$ one finds $H = - \dual F$, hence the first
equation becomes  $d \dual F = 0$.  Thus we have the vacuum Maxwell
equations.

\subsection{Translation invariance and Noether current}
Infinitesimal diffeomorphism invariance (in terms of the Lie
derivative) requires that (\ref{varL}) becomes an identity under the
replacement $\delta\to \Lie_N$:
\begin{equation}
d i_N {\cal L} \equiv \Lie_N {\cal L} \equiv d( \Lie_N \varphi
\wedge p) + \Lie_N \varphi \wedge \frac{\delta {\cal L}}{\delta
\varphi} + \frac{\delta {\cal L}}{\delta p} \wedge \Lie_N p.
\end{equation}
This simply means that ${\cal L}$ is a 4-form which depends on
position only through the fields $\varphi$, $p$. For this to be the
case the set of fields in ${\cal L}$ necessarily includes dynamic
geometric variables, which means gravity.

From this identity it follows that the ``translational current''
density (3-form)
\begin{equation}
{\cal H}(N) = \Lie_N \varphi \wedge p - i_N {\cal L}
\label{noethercur}
\end{equation}
satisfies the ``conservation law''
\begin{equation}
- d {\cal H}(N) \equiv \Lie_N \varphi \wedge \frac{\delta {\cal
L}}{\delta \varphi} + \frac{\delta {\cal L}}{\delta p} \wedge \Lie_N
p.
\end{equation}
Consequently, ``on shell'' (i.e.~when the field equations are
satisfied), the integral of the current over a spatial region will
give a conserved quantity for each vector field $N$. Note that, just
like other Noether conserved currents, ${\cal H}(N)$ is not unique:
it can be modified by adding the differential of any 2-form.

With geometric gravity included, we have also local diffeomorphism
invariance, which gives rise (in accordance with Noether's second
theorem) to a differential identity. Explicit calculation shows that
${\cal H}(N) = \Lie_N \varphi \wedge p - i_N {\cal L}$ always has
the form
\begin{equation}
{\cal H}(N) = N^\mu {\cal H}_\mu + d {\cal B}(N).\label{NH+dB}
\end{equation}
Thus we find that $d(N^\mu {\cal H}_\mu + d {\cal B}(N)) \equiv d
N^\mu \wedge {\cal H}_\mu + N^\mu d {\cal H}_\mu$  is proportional
to the field equations, therefore ${\cal H}_\mu$ vanishes ``on
shell''. Hence for gravitating systems the Noether translational
``charge''---energy-momentum---is quasi-local: it is given by the
integral of the boundary term, ${\cal B}(N)$. But this boundary term
as noted can be completely modified to any value. The Hamiltonian
approach includes an additional principle which naturally tames this
ambiguity.\cite{N04CQG}

\subsection{Hamiltonian approach}
Energy can be identified as the value of the Hamiltonian associated
with a time-like displacement vector field $N$. The Hamiltonian
$H(N)$ is given by an integral of a suitable Hamiltonian 3-form
(density) ${\cal H}(N)$ over a 3-dimensional (space-like) region
$\Sigma$. Generalizing $L = \dot q p - H$, from the first order
Lagrangian one constructs the Hamiltonian 3-form by projecting along
a ``time-like'' displacement vector field:
\begin{equation}
i_N {\cal L} = \Lie_N \varphi \wedge p - {\cal H}(N).
\end{equation}
The Hamiltonian density thus turns out to be just the Noether
translational current (\ref{noethercur}) identified above. As
already noted it satisfies the relation (\ref{NH+dB}) with ${\cal
H}_\mu$ vanishing on shell. Consequently the quasi-local
energy---regarded as the value of the Hamiltonian---is then
determined only by the boundary integral:
\begin{equation}
E(N) = \int_\Sigma {\cal H}(N) = \int_\Sigma \left[ N^\mu {\cal
H}_\mu + d {\cal B}(N) \right] = \oint_{\partial\Sigma} {\cal B}(N).
\end{equation}
The two parts of the Hamiltonian have distinct roles. The 3-form,
$\mathcal{H}_\mu$, although it has vanishing numerical value,
generates the equations of motion. For our concerns here the
Hamiltonian boundary term ${\cal B}(N)$ is the key quantity. It
plays a dual role: determining both the the quasi-local values and
the boundary conditions.

\subsection{Quasi-local quantities}
The Hamiltonian boundary term ${\cal B}(N)$ determines the various
quasi-local values corresponding to the Poincar\'e transformation of
space-time:
\begin{itemize}
\item
Energy $\longleftrightarrow$ $N$ a time-like displacement,

\item
Linear momentum $\longleftrightarrow$ a spatial translation,

\item
Angular momentum $\longleftrightarrow$ a rotation,

\item
center-of-mass moment $\longleftrightarrow$ a boost.
\end{itemize}
However we noted that ${\cal B}(N)$ can be adjusted; then it would
give different conserved values. What do they all these different
values mean physically?

\subsection{Boundary Conditions}
The variational principle contains an additional (largely
overlooked) feature which distinguishes all of these choices: {\it
the boundary variation principle}, i.e. the boundary term in the
variation tells us what to hold fixed on the boundary---it
determines the boundary conditions. Different Hamiltonian boundary
term choices are each associated with distinct boundary conditions.
(In this way this formalism gives a specific physical significance
to each of the traditional energy-momentum
complexes.\cite{CNC99PRL,CNC99,CN00GC})

This feature is similar to that of some familiar physical systems.
For example in thermodynamics the suitable measure of energy: the
internal energy, enthalpy, Helmholtz, or Gibbs free energy depends
on the system's boundary conditions. Another good example concerns
moving a dielectric within a parallel plate capacitor. The work
needed, and thus the appropriate energy density expression (the
symmetric or the canonical tensor) depends on the boundary
condition: fixed charge or fixed potential. Thus one can see that
there always are various distinct physical ``energies'' which
correspond to how a system interacts with the outside through its
boundary.

\subsection{Reference Configuration}
In general (in particular for gravity) it is necessary (technically,
in order to guarantee functional differentiability of the
Hamiltonian on the phase space with the desired boundary conditions)
to adjust the boundary term, ${\cal B}(N) = i_N \varphi \wedge p$,
which is naturally inherited from the Lagrangian. Moreover a {\it
reference configuration}, $\bar\varphi$ and $\bar p$, (which
determines the ground state) is essential (especially for gravity
where the ground state is not vanishing field but rather the
Minkowski metric) in particular to allow for the desired phase space
asymptotics.

\subsection{Quasi-local Expressions}
With $\Delta \varphi := \varphi - \bar\varphi$, $\Delta p := p -
\bar p$, where the bar indicates the reference value, we found two
boundary choices (essentially Dirichlet and Neumann) which have the
indicated covariant boundary terms in $\delta {\cal H}$.
\begin{eqnarray}
{\cal B}_{\varphi} &=& i_N \varphi \wedge \Delta p - \varsigma
\Delta \varphi \wedge i_N \bar p, \quad \to \quad i_N (\delta
\varphi \wedge \Delta p),
\\
{\cal B}_{p} &=& i_N \bar \varphi \wedge \Delta p - \varsigma \Delta
\varphi \wedge i_N p, \quad \to \quad - i_N (\Delta \varphi \wedge
\delta p).
\end{eqnarray}


\section{Application to GR}
For Einstein's (vacuum) gravity theory, General Relativity (GR) a
first order Lagrangian is
\begin{equation}
{\cal L}_{\rm GR} = \frac1{16 \pi} R^\alpha{}_\beta \wedge
\eta_\alpha{}^\beta,
\end{equation}
where $\Gamma^\alpha{}_\beta$ is the connection one-form,
$R^\alpha{}_\beta := d \Gamma^\alpha{}_\beta +
\Gamma^\alpha{}_\gamma \wedge \Gamma^\gamma{}_\beta$ is the
curvature 2-form and $\eta^{\alpha\beta} := *(\vartheta^\alpha
\wedge \vartheta^\beta)$.

Our general formalism with $\varphi \to \Gamma^\alpha{}_\beta$ and
$p \to \eta_\alpha{}^\beta$ gives the quasi-local expressions for
GR. We have two expressions for different types of boundary
conditions.   One of the choices stands out:\cite{CNT05PRD}
\begin{equation}
{\cal B}(N) := \frac1{16 \pi} \left[ \Delta \Gamma^\alpha{}_\beta
\wedge i_N \eta_\alpha{}^\beta + ({\bar D} N)_\beta{}^\alpha \Delta
\eta_\alpha{}^\beta \right].\label{favored}
\end{equation}
This is a Dirichlet type condition for a covariant object, the
orthonormal frame field. Asymptotically this expression gives not
only the ADM (spatial infinity) and Bondi energy (null infinty) but
also the Bondi energy flux. Moreover this expression is
distinguished by satisfying a positive energy property.

In the cases considered here, the contribution of the second term in
(\ref{favored}) vanishes.

\section{Homogenous Cosmologies}
Homogeneous cosmologies (non-isotropic in general) are described by
the Bianchi models:\cite{EM69CMP} the orthonormal coframe has the
form
\begin{equation}
\vartheta^0 = dt, \qquad \vartheta^a = h^a{}_k(t) \sigma^k,
\end{equation}
where the spatially homogeneous frames satisfy
\begin{equation}
d \sigma^k = \frac12 C^k{}_{ij} \sigma^i \wedge \sigma^j,
\end{equation}
where the $C^k{}_{ij}$ are certain constants.  The associated
space-time metric is thus
\begin{equation}
ds^2 = - dt^2 + g_{ij}(t) \sigma^i(x) \sigma^j(x),
\end{equation}
where $g_{ij} := \delta_{ab} h^a{}_i h^b{}_j$ (which need not be
diagonal).

There are 9 Bianchi types distinguished by the particular form of
the structure constants $C^k{}_{ij}$, especially by the value of
$A_k := C^i{}_{ki}$. They fall into two special classes:
\begin{itemize}
\item
{\it class A} ($A_k \equiv 0$): Types I, II, VI${}_0$, VII${}_0$,
VIII, IX;

\item
{\it class B} ($A_k \ne 0$): Types III, IV, V, VI${}_h$, VII${}_h$.
\end{itemize}
The respective scalar curvatures are: {\it vanishing} for Type I,
{\it positive} for Type IX, {\it negative} for all the other types.
It should be mentioned that certain special cases can be isotropic,
specifically isotropic Bianchi I, V, IX are equivalent to the usual
FRW $k = 0, -1, +1$.

For the natural choice of $N = \partial_t$, the  Dirichlet type
boundary condition, the Bianchi homogenous frame as boundary value,
and with the reference being the static homogenous cartesian frame,
the energy within a spatial volume $V$ according to our favored
quasi-local expression (\ref{favored}) is\cite{SNV07}
\begin{equation}
E(V) = \frac1{8\pi} A_j A_k g^{jk}(t) V(t) \ge 0.
\end{equation}
The result is true for all regions and for all types of sources
including dark matter, dark energy a/o a cosmological constant. More
specifically it vanishes for all class A models and is positive for
all class B models. Note: this is entirely consistent with the
important requirement that $E = 0$ for closed universes, since all
homogeneous class A models can be compactified and class B models
cannot.\cite{AS91CQG}

\section{FRW cosmology}
The Friedman-Robertson-Walker (FRW) (homogeneous {\it and}
isotropic) metrics have the form
\begin{equation}
ds^2 = - dt^2 + a^2(t) dl^2.
\end{equation}
The spatial metric $dl^2$ has constant curvature. The FRW spatial
metric has several equivalent manifestly
isotropic-about-a-chosen-point forms:
\begin{equation}
dl^2 = d \rho^2 + \Sigma^2 d\Omega^2 = \frac{dr^2}{1 - k r^2} + r^2
d\Omega^2 = \frac1{\left[ 1 + (k/4) R^2 \right]^2} \left( dR^2 + R^2
d\Omega^2 \right),
\end{equation}
where $\Sigma = (\sinh \rho, \rho, \sin \rho)$ for $k=(-1,0,+1)$,
respectively.

A natural choice in this case is $N = \partial_t$, Dirichlet type
boundary conditions, FRW frame boundary values, with the reference
being the static flat cartesian frame. The energies within a fixed
radius for the three FRW cases can be represented in several
equivalent forms (their identity follows from $\Sigma = r = R/(1 + k
R^2/4)$):
\begin{equation}
E = a \Sigma (1 - \Sigma') = a r \left[ 1 - \sqrt{1-k r^2} \right] =
\frac{a k R^3}{[1 + (k/4) R^2]^2}.
\end{equation}

More specifically,
\begin{eqnarray}
E_{k = -1} &=& a \sinh\rho (1 - \cosh\rho) = a r \left[ 1 - \sqrt{1
+ r^2} \right] = \frac{- a R^3}{2 (1 - R^2/4)^2} \quad \le 0,
\nonumber\\
E_{k = 0} &=& 0,
\nonumber\\
E_{k = +1} &=& a \sin\rho (1 - \cos\rho) = a r \left[ 1 - \sqrt{1 -
r^2} \right] =  \frac{a R^3}{2 (1 + R^2/4)^2} \quad \ge 0.
\end{eqnarray}

\section{Discussion}
According to our favored quasi-local energy expression, homogeneous
choices give vanishing energy for all regions of Bianchi class A
models and positive energy for class B. Isotropic choices give
energies proportional to the spatial curvature parameter $k$:
vanishing for flat, {\it negative} for the open model, and positive
for the closed model (but nevertheless vanishing, as required, when
the considered volume is extended to include the whole universe).


Some of the Bianchi models can be isotropic, specifically
\begin{itemize}
\item
isotropic Bianchi I (class A) $\equiv$ FRW${}_{k = 0}$;

\item
isotropic Bianchi IX (class A) $\equiv$ FRW${}_{k = +1}$;

\item
isotropic Bianchi V \& VII${}_h$ (class B) $\equiv$ FRW${}_{k =
-1}$.
\end{itemize}
Note that our quasi-local expression thus can give different energy
values to exactly the same geometry.  This is not at all mysterious;
it is clearly a consequence of different reference and boundary
value choices.  Homogenous boundary values are not the same as
isotropic boundary values.  To understand the physical and geometric
meaning of the differences between the homogeneous and isotropic
choices  in detail, we need to do more calculations using the rather
complicated relations between the FRW and Bianchi coordinates.
Meanwhile from our analysis it seems that the homogeneous choice is
more suitable physically than the isotropic-about-a-chosen-point
choice in general, since it gives a non-negative energy.

It is also noteworthy that for the case of the open FRW ($k = - 1$)
with vanishing matter, the solution to the Einstein equation gives
$a(t) = t$. It can be directly verified that the geometry is then
really just Minkowski space in non-standard coordinates, yet our
expression gives a {\it non-vanishing energy}, which, moreover is
{\it negative} with the FRW choices.

Concerning two of the quasi-local desiderata, for the expression and
boundary/reference choices considered we found that
\begin{itemize}
\item
positivity need not hold;

\item
``zero energy iff flat Minkowski space'' need not hold in either
direction.
\end{itemize}
It seems that these quasi-local criteria should be reconsidered.

As we prefer positivity, we are inclined to see our negative result
as disfavoring the FRW isotropic-about-a-chosen-point
boundary/reference choice.

Regarding $E = 0$, clearly any {\it homogeneous} measure of
quasi-local energy in Bianchi I models must necessarily vanish for
all regions---since these models can be compactified (with 3 torus
topology identifications) on any scale, and the energy of a closed
universe must vanish.  In light of this, we propose that the
``unique quasi-local $E = 0$ Minkowski ground state'' requirement be
replaced by something like ``$E(V) = 0$ iff a neighborhood of $V$
can be compactified.''

\section*{Acknowledgments}
This work was supported by the National Science Council of the
R.O.C. under the grants NSC 95-2112-M-008-003 (CMC) and NSC
95-2119-M-008-027 (JMN). CMC and JMN were supported in part by
National Center of Theoretical Sciences and the (NCU) Center for
Mathematics and Theoretical Physics.



\begin{thebibliography}{0}
\bibitem{CNC99PRL}
  C.-C. Chang, J. M. Nester and C.-M. Chen,
  ``Pseudotensors and quasilocal energy-momentum,''
  {\it Phys. Rev. Lett.} {\bf 83}, 1897-1901 (1999) [arXiv:gr-qc/9809040].

\bibitem{N04CQG}
  J. M. Nester, ``General
pseudotensors and quasilocal quantities'',
  {\it Class.~Quantum Grav.} {\bf 21}, S261-S280 (2004).

\bibitem{MTW} C.~W. Misner, K.~Thorne and J.~A. Wheeler, {\it Gravitation},
(Freeman, San Francisco, 1973).

\bibitem{S04LRR}
  L.~B.~Szabados,
  ``Quasilocal energy-momentum and angular momentum in GR: a review article,''
    {\it Living Rev. Relativity,} {\bf 7}, 4 (2004).
\\
{\tt http://www.livingreviews.org/lrr-2004-4}.

\bibitem{LY04}
  C.~C.~Liu and S.~T.~Yau,
  ``Positivity of quasi-local mass II,''
  arXiv:math.dg/0412292.


\bibitem{CNT95PLA}
  C.-M. Chen, J. M. Nester, and R.-S. Tung,
  ``Quasilocal energy-momentum for geometric gravity theories,''
  {\it Phys. Lett. A} {\bf 203}, 5-11 (1995) [arXiv:gr-qc/9411048].

\bibitem{CN99}
  C.-M. Chen and J. M. Nester,
  ``Quasilocal quantities for GR and other gravity theories,''
  {\it Class.~Quantum Grav.} {\bf 16}, 1279-1304 (1999) [arXiv:gr-qc/9809020].


\bibitem{CNC99}
  C.-C. Chang, J. M. Nester, and C.-M. Chen,
  ``Energy-momentum (quasi-)localization for gravitating systems,''
  in {\it Gravitation and Astrophysics} ed Liao Liu, Jun Luo, X.-Z. Li, J.~P. Hsu
  (World Scientific, Singapore) pp 163-73 [arXiv:gr-qc/9912058].

\bibitem{CN00GC}
  C.-M. Chen and J. M. Nester,
  ``A symplectic Hamiltonian derivation of quasilocal energy-momentum for GR,''
  {\it Gravitation \& Cosmology} {\bf 6}, 257-270 (2000) [arXiv:gr-qc/0001088].

\bibitem{CNT05PRD}
  C.-M. Chen, J. M. Nester and R.-S. Tung,
  ``The Hamiltonian boundary term and quasilocal energy flux,''
  {\it Phys. Rev.} {\bf D72}, 104020 (2005) [arXiv:gr-qc/0508026].

\bibitem{EM69CMP}
  G.~F.~R.~Ellis and M.~A.~H.~MacCallum,
  {\it Comm. Math. Phys. \bf 12} (1969) 108--141.

\bibitem{SNV07}
  L.~L.~So, J.~M.~Nester, and T.~Vargas,
  ``On the energy of homogeneous cosmologies'', in preparation.

\bibitem{AS91CQG}
  A. Ashetkar and J. Samuel,
  {\it Class. Quantum Grav. \bf 8} (1991) 2191--2215.

\end{thebibliography}
\end{document}